# Mobile Location Update using Distance Method

*Mohammad Wasif*
HCL Technologies Ltd, Chennai, Tamil Nadu, India – 600096

***Abstract:*** *General Packet Radio Service (GPRS) is a complex data network which upgrades current second generation GSM networks, offering true high-speed internet (IP) and network connectivity over existing GSM cellular networks. The increasing population of mobile users leads to congestion problems in these systems, and motivates the development of more efficient management schemes. This project deals with radio resource and mobility management such as location management and handoff management using distance method in GPRS networks. A simulator based on MATLAB which can study the location updating is used in this GPRS system.*

## 1. INTRODUCTION

As GSM is already the most broadly deployed digital wireless standard in the world, with over 850 million users in over 195 countries and with service provided by over 400 operators, GSM represents over 70% of today's digital wireless market. The infrastructure and technology for connecting mobile devices, for global deployment and the billing arrangements and tariff already exist making it an ideal platform for Machine to Machine applications (M2M). GPRS now makes it possible to deploy several new devices that have previously not been suitable over traditional GSM networks due to the limitations in speed (9600bps), message length of the Short Message Service (160 characters), dial up time and costs[2]. Packet-switched data under GPRS is achieved by allocating unused cell bandwidth to transmit data. As dedicated voice (or data) channels are setup by phones, the bandwidth available for packet switched data shrinks. A consequence of this is that packet switched data has a poor bit rate in busy cells. The theoretical limit for packet switched data is 171.2 kbits/s. A realistic bit rate is 30–80 kbits/s, because it is possible to use max 4 time slots for downlink. These applications include Point Of Sale Terminals, Vehicle tracking systems, and monitoring equipment. It's even possible to remotely access and control in-house appliances and machines.

GPRS is different from the Circuit Switched Data (or CSD) connection included in GSM standards. In CSD, a data connection establishes a circuit, and reserves the full bandwidth of that circuit during the lifetime of the connection. GPRS is packet-switched which means that multiple users share the same transmission channel, only transmitting when they have data to send. This means that the total available bandwidth can be immediately dedicated to those users who are actually sending at any given moment, providing higher utilization where users only send or receive data intermittently. Web browsing, receiving e-mails as they arrive and instant messaging are examples of uses that require intermittent data transfers, which benefit from sharing the available bandwidth.

The highly dynamic and busty nature of packet switched services such as GPRS will require a much more flexible method of radio resource management so as to maximize system resources[1]. In order to integrate GPRS to support GSM architecture, a new class of network nodes, called GPRS support nodes (GSN) and serving GPRS support node (SGSN) has been introduced as shown in Figure 1. GSN are responsible for the delivery and routing of data packets between the mobile stations and the external packet data networks. Radio Resource is to establish, maintain and release communication links between mobile stations and MSC. Mobility management is used to track mobile location within each mobile network. If the mobile terminal updates its location whenever it crosses a cell boundary, the network can maintain its precise location, thus obviating the need for paging[1]. Current personal communication uses Location Area based update algorithm. The coverage area is partitioned into a number of LA's each containing a group of cells. This increases the paging cost.

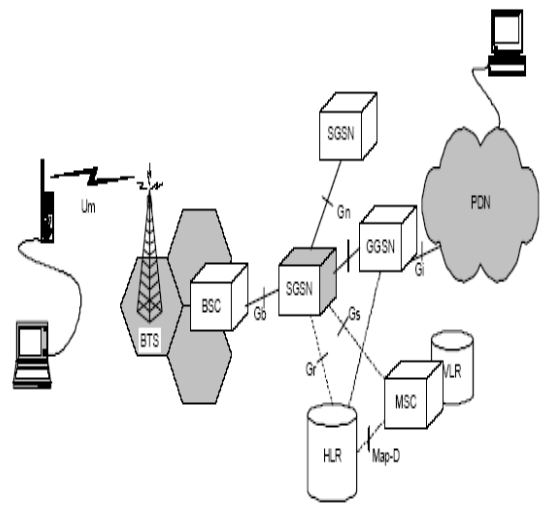

Fig 1: Logical architecture of GPRS network

The goal of distance based location update scheme is to reduce the update cost by taking advantage of user's mobility pattern. As long as the mobile moves in the same area, location update is not necessary. The update will take place only when mobile moves to new location area. GPRS achieves faster connection speeds thanks to two cutting-edge technologies. The first is the General Packet. Rather than sending information in a steady stream through a single channel as current phones do, a GPRS enabled device breaks the information down into "packets" and sends them over multiple channels (up to eight). Each packet travels by the quickest available route to the recipient, where it is reassembled into the original message. Sending packets by several different channels increases the speed of transmission and cuts down on signal errors.

## 2. Radio Resource and Mobility Management

Radio Resource function is to establish, maintain and release communication links between mobile stations and MSC. The elements that are mainly concerned with the RR function are the mobile station and the base station. However, as the RR function is also in charge of handovers, is also concerned with the RR functions. Radio Resources Management Controls the setup, maintenance, and termination of radio and fixed channels including handovers.

Mobility Management manages the location updating and registration procedures as well as security and authentication[2]. Mobility management functions are used to track its location within each mobile network. The SGSN's communicate with each other and update the user location. The MS profiles are preserved in the Visitor Location Registers (VLR's) that are accessible by the SGSN in each mobile network. At the end of transmission or when a MS moves out of the area of a specific SGSN, the logical link is released and the resources associated with it can be reallocated.

Location management is the means by which the GPRS network keeps track of the MS location[7]. Within a GPRS network, three types of location management procedures are described:
- *Cell update* is the means by which a MS informs the network of its current cell location.
- *Intra-SGSN routing* update is the procedure used when a MS changes RA and remains serviced by the same SGSN.
- *Inter-SGSN routing* update is the procedure used when the entry of a MS to a new RA triggers a change of SGSN service area.

The location update determines when or where an MT should report its location to the network. In the Static Schemes a location update is performed when a MT changes its LA. It cannot be adjusted based on the parameters of a MT from time to time compared to the Dynamic Schemes where it can be adjusted based on the parameters of a MT from time to time. Most of the recent research focuses on the dynamic schemes. Other update schemes are selective LA update scheme, profile-based location update scheme, timer-based update scheme, movement based update scheme and LeZi update.

## 3. Limitations of GPRS

GPRS transmission rates are much lower than in theory. To achieve the theoretical maximum of about 170kbit/s would require allocating eight time slots to a single user which is not likely to be allowed by network operators. Even if this maximum allocation was allowed, the GPRS terminals may be constrained by the number of time slots they can handle. GPRS relies on packet switching which means that data packets can traverse different routes and then be reassembled in their final destination leading to potential transit delays affecting the Quality of Service. GPRS relies also on re-transmission and data integrity protocols to ensure that data packets transmitted over the radio air interface are not lost or corrupted. This can affect even further the transit delay problem. GPRS allows the specification of QoS profiles using service precedence, delay, reliability, mean and peak throughput.

Although these attributes are signaled in the protocols and are negotiated between the network and the MS, no procedures are defined to provide QoS differentiation between services. This causes a lack of uniformity with respect to QoS between manufacturers and potentially between operators. This issue is being addressed in later standard efforts. Although it is possible theoretically to specify a high QoS profile, traffic over radio imposes severe constraints on the Quality of Service[7]. The protocols between the BSS and the SGSN support mainly asynchronous data transfer applications making it a challenge to implement real-time interactive traffic.

## 4. Present Location Update Scheme

Current personal communication network uses a location area (LA)-based update algorithm. The coverage area is partitioned into a number of LA's each containing a group of cells. All base stations within the same LA broadcast the identifier (ID) of their LA periodically. Each mobile terminal compares its registered LA ID with the current broadcast LA ID. Location update is triggered if the two ID's are different. Upon a call arrival for a particular mobile terminal, all the cells within its current LA are polled simultaneously, ensuring success within a single step. The major drawback of this method is that since the number of cells within a typical LA is large, the paging cost is very high.

Location management is concerned with the procedures required to enable the network to maintain the location information for each subscriber, or more specifically, for each active mobile terminal with a registered subscriber and to efficiently handle the establishment of incoming calls. The location management procedures are required in cellular networks to manage subscriber mobility. One of the fundamental properties of cellular network is that, because subscribers can roam throughout the network and possibly in other networks, there can no permanent connection to subscriber as is the case in fixed telephony. In order to be able to route incoming calls to mobile subscribers, the network needs the current location of each subscriber.

A mobile station communicates with another station, either mobile or land, via a base station. A mobile station can not communicate with another mobile station directly. To make a call from a mobile station, the mobile station first needs to make a request using a reverse control channel of the current cell. If the request is granted by MSC, a pair of voice channels will be assigned for the call. To route a call to a mobile station is more complicated. The network first needs to know the MSC and the cell in which the mobile station is currently located. How to find out the current residing cell of a mobile station is an issue of location management. Once the MSC knows the cell of the mobile station, the MSC can assign a pair of voice channels in that cell for the call. If a call is in progress when the mobile station moves into a neighbor cell, the mobile station needs to get a new pair of voice channels in the neighbor cell from the MSC so the call can continue. This process is called handoff (or handover). The MSC usually adopts a channel assignment strategy which prioritizes handoff calls over new calls. Vertical handoff is the first level of handoff granularity, in which handoff occurs between access points that are using different access network technologies. It is a crucial enabling technology for heterogeneous networks. Typical examples could be e.g. handoff between LAN and WLAN, and between WLAN and GPRS. Horizontal handoff is the handoff between access points of the same type of network technology. This is the traditional definition of handoff for homogeneous network systems. Horizontal handoff occurs with different granularities. Macro-handoff occurs between different visited domains, while micro-handoff occurs between different adjacent location areas, access point regions, cell zones, or even logical channels in one cell.

### 4.1. States

The GPRS MS can share radio channels with other subscribers connected to the network. For this reason, the MS is defined to have three possible states: IDLE, READY and STANDBY.

In **IDLE** state or mode, the subscriber is not attached to the GPRS MM. There is no valid location or routing information for the subscriber. Thus the GPRS can not reach the MS. However the MS does the selection of PLMN and GPRS cell selection and cell handover.

In order to get to the **READY** state the MS performs the attachment and MS's subscription information is downloaded from HLR to SGSN if it is the first attachment. The MSC/VLR is also updated in case of International Mobile Subscriber Identity (IMSI) attachment is also performed. The MS is able to send and receive data packet normally. The network knows the cell, where the subscriber is located. Cell update is performed, when MS is doing a cell handover at the same Routing Area (RA). In the case when RA is also a new, the RA update is performed. Routing area is a subset of a Location Area (LA), so a LA update is possible. In MS initiated detach, the MS informs about it to the SGSN.

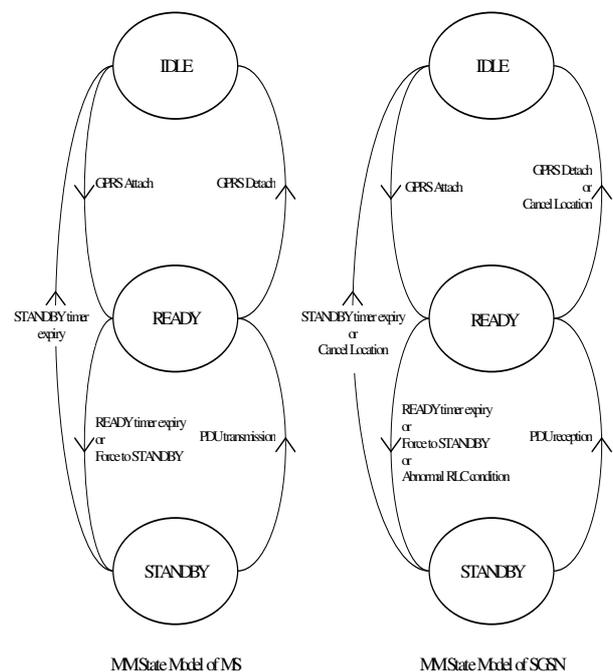

Fig 2: Functional Mobility Management state model for MS and SGSN in non-anonymous access.

In **STANBY** state, the MS is still attached to GPRS MM but PTP transmission is not possible. The MS can initiate activation or deactivation of PDP context to start the packet transmission. The MS sends only RA information in case of a RA handover. After the timer expires, the MM context can be deleted in IDLE state. The three states are shown in Figure 2.

### 4.2. Distance method

The goal of distance based location update scheme is to reduce the update cost by taking advantage of the user's mobility pattern. The distance here is defined in term of cells. When the distance reaches a predefined threshold say D, the mobile station updates it location. The distance based strategy guarantees that the mobile station is located in an area that is within a distance from the last reported cell. This area is called the residing area of the mobile station. Whenever an incoming call arrives for a mobile station, the cellular system will page all the cells within a distance of, D from the last reported cell. The distance based strategy is dynamic, and the distance threshold, D can be determined on a per-user basis depending on his/her mobility pattern. Update will take place only when mobile moves to new location area.

The network maintains a profile for each user, which includes a sequential list of the LA's the user is most likely to be located at in different time periods. This list is sorted from the most to least likely LA where a user can be found. When a call arrives, the LA's on the list are paged sequentially. As long as the mobile terminal moves between same location area, no location update is necessary. The update will take place only when the mobile moves to the new location area. The present scheme treats according to the size of cell versus speed of mobile. When the mobile move to the new location area crossing the cell boundary, the update will be calculate. More update will take place if the size of the cell is small compare with less update for large size of cell by different amount of velocity.

In these schemes each user registers to its corresponding HLR. This scheme maintains one update database having the following scheme: user ID, user velocity, user direction, cell ID and time. The MSC/VLR keeps track of a set of MS movement. Whenever there is any change of movement which consists of user velocity and direction, the current profile of user is sent to HLR and count the update. Other update schemes are selective LA update scheme, profile-based location update scheme, timer-based update scheme, movement based update scheme and LeZi update.

**4.3. Simulation and Results**

The location update simulation is done in MATLAB for different velocities (maximum speed). The 28 hexagonal shaped cells are provided with a base station (BS) each at the centre for omni directional coverage. All users (MS) moves with same velocity and also they move randomly. The size of hexagonal cell is same for all the cells. In such a cellular configuration, user moves across cells, crossing cells boundaries impacting on the quality of received signal of their resident cell.

The MS is considered to be in a particular cell if it is only nearer to the base station (BS) of that particular cell in which it is moving. In the case of **IDLE** mode, there is no need of location update since the mobile will be out of the coverage area of the base station, but in the case of **READY** and **STANDBY** modes, the update is done.

Performance results of update are shown in fig.3 and fig.4. In fig.3, for the number of stations= 10, range of radio=2, time scale of random motion=100, average transmission time of packets=12, simulation time =30000ms. The update values are

update for each cell = 0   2   5   8   0   0   0   8   10   5

total update = 38

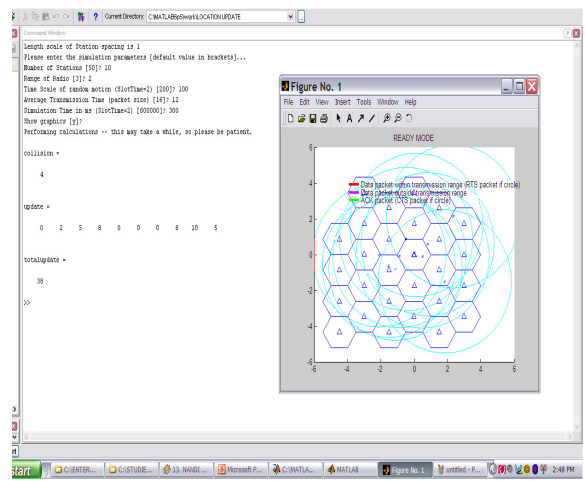

Fig. 3: Simulated output

In fig.4, for the number of stations=20, range of radio=2, time scale of random motion=200, average transmission time of packets=12, simulation time =60000ms. The update values are

update = 0   0   0   0   2   9   8   19   0   0   18   17   0   4   15   26   21   0   13   10

Total update = 162

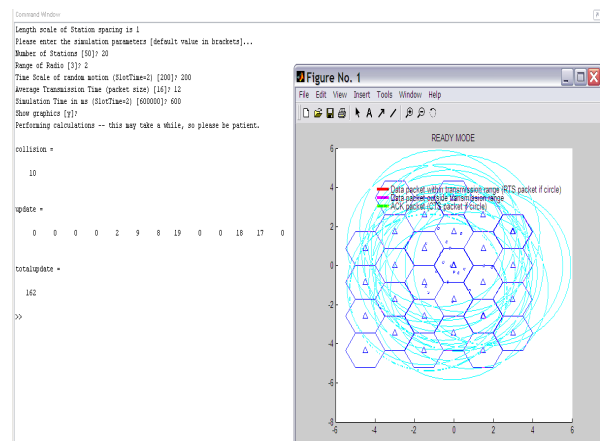

Fig. 4: Simulated output

The graphs for HLR update values obtained for different user velocities and for the number of users, n=10, n=20 and n=30 are shown in the figures 5, 6 and 7 respectively.

The update will be more in case of more users. As in fig.5 for the number of users n=10, the user update values are less compared to n=20 in the case of fig. 6 for different velocities. Also there is more update in fig. 7 where n=30 compared to fig. 6 for different velocities. In all the graphs the updating of mobile users are almost similar for both READY and STANDBY states. If the size of the cell is varied then there will be an increase in update also. Since the movement of the mobile users will be more random than in the case of large cells the values of update will be more. The READY mode values are plotted in the graphs with a blue colored line and the red line for STANDBY mode values. The graph is drawn for velocity (Max speed) and HLR updates.

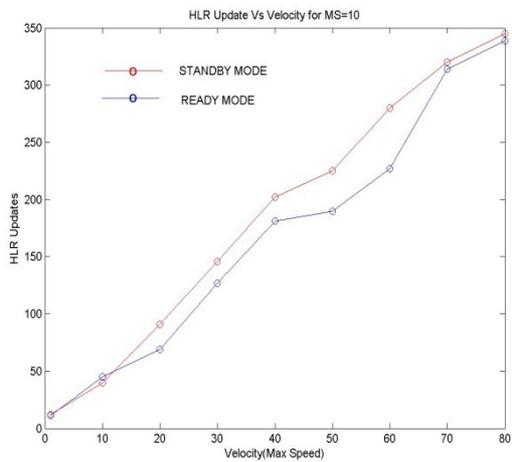

Fig 5. HLR vs velocity for MS=10

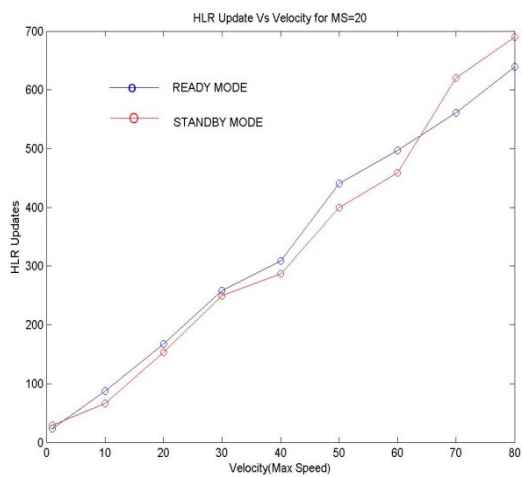

Fig 6. HLR vs velocity for MS=20

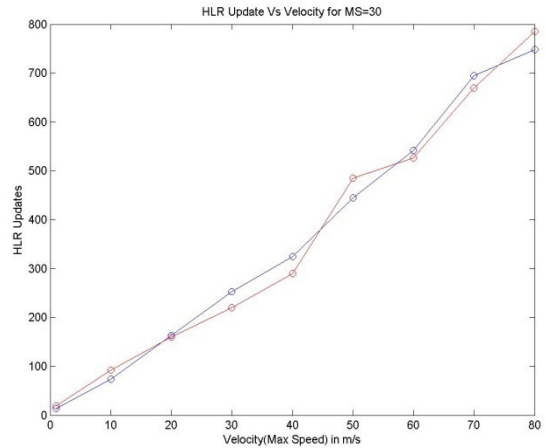

Fig 7. HLR vs velocity for MS=30

These results are examined and compared with other methods like selective LA update scheme, profile-based location update scheme, timer-based update scheme, movement based update scheme for further advancements. For future methods like predictive distance based location update scheme which is the extension of distance based location update these graphs and results will be useful to analyze and obtain better results.

Also the security concerns play an important role in mobile networks and gained a special attention in the GSM world. GPRS provides a security function similar to that of GSM. It is responsible for authentication and service request validation to prevent unauthorized service usage. User confidentiality is also protected using temporary identification during connections to the GPRS network. Finally, user data is protected using ciphering techniques. These security concerns are also should be considered for further proceedings.

## 5. Conclusion

There is no straightforward solution that takes account of the multiplicity of location management requirements. The distance method reduces the update cost since update is needed only when the mobile moves to new location area. GPRS has a better use of radio and network resources and completely transparent IP support. GPRS networks will have to be tuned and optimised to the point that they provide acceptable levels of service quality. In this paper, the update is done for mobiles with different velocities using distance method and the graphs are drawn for different modes and analysed. Also methods like predictive-distance based management, profile-based management are proposed to achieve advanced location management towards a complete solution. Finally there should be an increase in traffic volumes and the bottlenecks should be removed which will benefit customers.

## Author Introduction

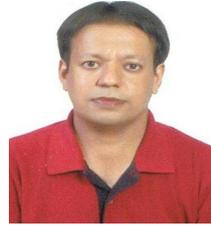

**Mohammed Wasif** is a Technical Leader in HCL Technologies Ltd, India and is currently leading a project for Juniper Networks Ltd. Before joining HCL Technologies Ltd, he worked as an IT Analyst in Tata Consultancy Services Ltd, India. He completed his Master of Technology (M.Tech) in Communication Engineering from VIT University, Vellore, India in 2004 and did his Bachelor of Engineering (B.E) in Electronics and Communication Engineering discipline at M.J.P Rohilkhand University, Bareilly, India in 2002. He has published Research papers in peer-reviewed international journal and in IEEE international conferences and his research interests include VLSI for Wireless Communication Systems, MIMO Channel, Network-On-Chip (NOC), SDR, VOIP, IMS, Mobile Ad-Hoc and High-Performance Computer Networks.